\documentclass[journal]{IEEEtran}
\usepackage{cite}
\usepackage[cmex10]{amsmath}
\usepackage{amssymb}
\usepackage{graphicx}
\usepackage{epstopdf}
\usepackage{graphics}
\usepackage{pdfpages}
\usepackage{float}
\usepackage{changes}
\usepackage{multirow}

\ifCLASSINFOpdf
\else
\fi
\usepackage[cmex10]{amsmath}
\usepackage{epsfig} 
\ifCLASSOPTIONcompsoc
  \usepackage[caption=false,font=normalsize,labelfont=sf,textfont=sf]{subfig}
\else
  \usepackage[caption=false,font=footnotesize]{subfig}
\fi
\usepackage{stfloats}
\begin{document}


\title{OFDM With Hybrid Number and Index Modulation}
\author{Ahmad M. Jaradat, Jehad M. Hamamreh,
        and Huseyin Arslan,~\IEEEmembership{Fellow,~IEEE}
\thanks{A. M. Jaradat is with Department of Electrical and Electronics Engineering, Istanbul Medipol University, Istanbul, 34810, Turkey (email: ajaradat@st.medipol.edu.tr,jaradat@ymail.com). J. M. Hamamreh is with the Department of Electrical and Electronics Engineering, Antalya Bilim University, 07468 Antalya, Turkey (email: jehad.hamamreh@gmail.com,jehad.hamamreh@antalya.edu.tr).}
\thanks{H\"{u}seyin Arslan is with Department of Electrical and Electronics Engineering, Istanbul Medipol University, Istanbul, 34810, Turkey and also with Department of Electrical Engineering, University of South Florida, Tampa, FL, 33620, USA (e-mail: huseyinarslan@medipol.edu.tr).}}

\maketitle
\begin{abstract} 
A novel transmission scheme is introduced for efficient data transmission by conveying additional information bits through jointly changing the index and number of active subcarriers within each orthogonal frequency division multiplexing (OFDM) subblock. The proposed scheme is different from the conventional OFDM-subcarrier number modulation (OFDM-SNM) and OFDM-index modulation (OFDM-IM), in which data bits are transmitted using either number or index of active subcarriers. The proposed modulation technique offers superior spectral and {energy} efficiency compared to its counterparts OFDM-SNM and OFDM-IM, especially at low modulation orders such as binary phase shift keying (BPSK) that can provide high reliability and low complexity, making it suitable for Internet of Things (IoT) applications that require better spectral and {energy} efficiency while enjoying high reliability and low complexity. Bit error rate (BER) performance analysis is provided for the proposed scheme, and Monte Carlo simulations are presented to prove the consistency of simulated BER with the analyzed one.

\end{abstract} 

%
\IEEEpeerreviewmaketitle
\section{Introduction} %

 \IEEEPARstart{W}{ide} scope of use cases along with various demands and features should be handled by 5G and beyond networks \cite{6824752,6736749}. The future networks should have the capabilities of supporting different, contrasting requirements in terms of ultra-high reliability, very low latency, very high data rates, improved energy efficiency (EE), and low computational complexity \cite{Zaidi_5G_2016,Ankarali2017}. Thus, it is very critical to define an appropriate modulation for a specific 5G use case \cite{6923528,8085125,Jaradat2019}. 
 
 
Multicarrier transmission schemes are heavily used in wireless communications. The fact behind this extensive employment of multicarrier techniques is their attractive features like easy equalization, multi-user scheduling, and support for adaptive modulation and coding techniques, etc \cite{Sahin2014}. The OFDM transmission scheme is considered as one of the interesting multicarrier techniques. The OFDM scheme is employed in broad range of applications, standards, and communication systems \cite{Hwang2009}. 

However, the conventional OFDM scheme has many limitations with respect to a set of performance metrics \cite{vaezi2018multiple}, which opens the door for multiple options of OFDM-based improved modulation schemes with respect to some performance evaluation metrics such as reliability, {spectral efficiency (SE)}, {EE}, etc\cite{Jaradat2019}. {Improving SE and EE of modulation options becomes one of the major issues for the future wireless communications. Particularly, achieving significant EE is one important criteria to employ a modulation scheme under the millimeter wave band \cite{8663599}.} 

Different OFDM-based modulation options have been proposed in the literature and their classification, comparison and future directions have been provided in \cite{Jaradat2019}. Among the promising OFDM-based modulation options, index and number-based OFDM modulation schemes are attractive transmission techniques where some subcarriers are conventionally modulated alongside the additional bits transmitted by exploiting the index or number dimension \cite{Jaradat2019}. The extra degrees of freedom offered by such promising modulation schemes are exploited to enhance the performance of conventional OFDM in terms of some performance metrics such as throughput, reliability, {EE}, etc.
 \;{In particular}, OFDM-index modulation (OFDM-IM) \cite{Basar_trans_2013} and OFDM-subcarrier number modulation (OFDM-SNM) \cite{Jaradat2018} are among the promising index, and number-based modulation techniques for OFDM-based waveforms, respectively \cite{Jaradat2019}. 

The OFDM-IM technique exploits specific active subcarrier indices in order to transmit additional information, but this fixed activation ratio per subblock limits the enhancement of SE for the conventional OFDM-IM scheme. The design of OFDM-IM {subcarrier activation method (i.e. the codebook)} depends on either a look-up table or the combinatorial method which is of high complexity and has not fully utilized the frequency selectivity for reliability improvement \cite{GokceliAccess2017}. Several improved OFDM-IM schemes have been proposed in \cite{IMmag2016,wen2016index,Basar_access_2017,Bharath2017,Mao_survey_2018,Cheng5G2018} with the main motivation of enhancing the SE of the conventional OFDM-IM scheme.

 Among these improved OFDM-IM schemes, a generalized version of OFDM-IM has been developed known as OFDM with generalized index modulation (OFDM-GIM) \cite{Fan2015}. OFDM-GIM is very much similar in principle to the basic OFDM-IM (where information bits are sent by indices) with the only difference that instead of using just one fixed activation ratio throughout the whole OFDM block, different activation ratios are used per each subblock within the whole OFDM block in a deterministic, fixed manner (not random based on the incoming data). This results in changing the number of active subcarriers (but there are no bits sent by the number of subcarriers, where the information bits are still sent by indices). Moreover, the selection of the activation ratios in OFDM-GIM must be shared with the receiver so that it can perform the detection of the active subcarrier indices.

Some improved OFDM-IM schemes such as dual-mode index modulation aided OFDM (DM-OFDM) \cite{MaoHanzo2017} and its generalized version called generalized DM-OFDM (GDM-OFDM) \cite{GDM2017} as well as OFDM with multi-mode IM (OFDM-MMIM) \cite{Wen2017} overcome the main limitation of OFDM-IM by transmitting data symbols on all available subcarriers along with IM. Aside from the enhancement in SE provided by these improved OFDM-IM schemes, it is difficult to mitigate intercarrier interference (ICI) and/or reduce peak-to-average power ratio (PAPR) due to activating all OFDM subcarriers.

Based on the general shape of the OFDM-IM block, a new number-based OFDM transmission scheme called OFDM-SNM is proposed in \cite{Jaradat2018}. This novel modulation scheme implicitly conveys information by utilizing numbers, instead of indices, of turned on subcarriers alongside the conventional symbols. 
An enhanced scheme of OFDM-SNM is proposed in \cite{8703169}, which exploits the flexibility enabled by the original OFDM-SNM scheme \cite{8703169} by placing the active subcarriers adaptively based on the channel to offer an additional coding gain in the high signal-to-noise ratio (SNR) region. The adaptation is performed based on the instantaneous channel state information {(CSI)} in which the incoming information bits are dynamically mapped to subcarriers with high channel power gains. 

Inspired by these unique features of the OFDM-IM and OFDM-SNM schemes, we introduce a novel alternative transmission scheme called OFDM with hybrid number and index modulation (OFDM-HNIM) which is based on the combination of both OFDM-SNM and OFDM-IM schemes. \textbf{Our main contributions} are summarized as follows:
\begin{itemize}
  \item We develop and propose a novel modulation scheme called OFDM-HNIM in which the available subcarriers are partitioned into subblocks, and information bits are conveyed not only by the modulated subcarriers but implicitly also by both numbers and indices of active subcarriers relying on the incoming bits in order to embed extra information besides the $M$-ary constellation symbols. 
  Therefore, the SE is significantly improved compared to OFDM-SNM and OFDM-IM schemes while maintaining low detection complexity.
  \item We derive tight, closed-form expressions for upper bound on the BER of OFDM-HNIM systems assuming ML detection. Two different ML detectors have been employed in the proposed OFDM-HNIM scheme: Perfect subcarrier activation pattern estimation (PSAPE), and imperfect SAP estimation (ISAPE) ML-based detectors. {Moreover, log-likelihood ratio (LLR) detector is employed on the proposed OFDM-HNIM scheme}. The BER performance of OFDM-HNIM is further compared with that of the existing state-of-the-art including OFDM-IM, OFDM-SNM, {and conventional} OFDM through Monte Carlo simulations.
\end{itemize}
\textbf{The merits} of the proposed scheme can be stated as follows:
\begin{itemize}
    \item The proposed OFDM-HNIM scheme increases the system design flexibility by creating an extra degree of freedom in the number and index dimensions, which can be exploited for different purposes such as enhancing the overall SE of the communication system.
    
    \item Unlike conventional OFDM, DM-OFDM, and OFDM-MMIM where all subcarriers are occupied by non-zero elements, exploiting the inactive subcarriers featured in the proposed OFDM-HNIM scheme could be used to lessen interference among subcarriers, and enhance {EE} by reducing PAPR. {Moreover,} it is possible to utilize the inherent features of OFDM-SNM in placing the active subcarriers in positions where the resultant system provides minimum levels of ICI and/or PAPR. 
    
    \item Individual SNM and IM modules could be exploited for the applications where high throughput or {EE} are needed, respectively. By utilizing the hybrid mapper of SNM and IM modules, high throughput alongside {EE} could be obtained for some applications.
\end{itemize}

The remainder of this paper is prepared as follows. The proposed OFDM hybrid scheme with number and index modulation scheme is explained in Section II. Performance evaluation of the OFDM-HNIM scheme in terms of SE, average bit error probability (ABEP), {EE}, and computational complexity is given in Section III. In Section IV, the simulation results and comparisons between the proposed hybrid scheme and its competitive schemes are presented. Finally, Section V presents concluding remarks. 

\textit{Notation}: Matrices and column vectors are represented by bold, capital and lowercase letters, respectively. $\circledast$, $E(.)$, $(.)^{T}$, $(.)^{H}$, and $|.|$ represent circular convolution operation, expectation, transposition, Hermitian transposition, and absolute value, respectively. det($\mathbf{A}$) denotes the determinant of $\mathbf{A}$. ${n\choose k}=\frac{n!}{k! (n-k)!}$ represents the binomial coefficient. $\mathbf{A}\sim \mathit{\cal CN}(\mu,\sigma^2)$ represents the complex Gaussian distribution of $\mathbf{A}$ with mean and variance of $\mu$ and $\sigma^2$, respectively. $Q(.)$ represents the Q-function. $\sim \mathcal{O}(.)$ denotes the complexity order of a technique. $\mathbf{H}(\mathbf{A})$ represents the entropy of $\mathbf{A}$, and $\mathbf{I}(\mathbf{A})$ represents the mutual information of $\mathbf{A}$.
\section{Proposed Hybrid scheme} 
\begin{figure}[t]
\centering
\includegraphics[height=2.3in,width=3.4in]{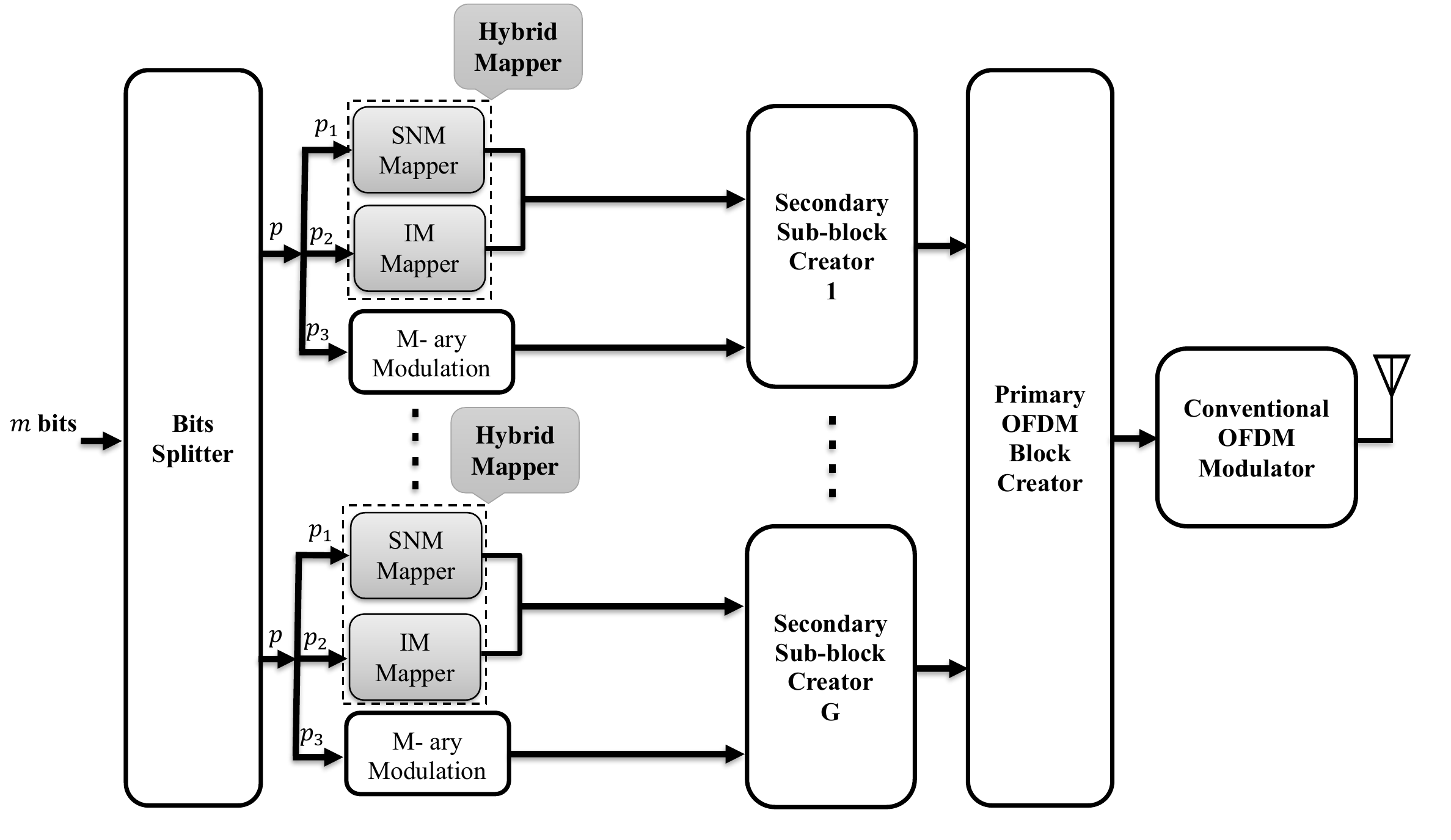} 

\caption{{The proposed OFDM-HNIM} transmitter structure.}
\label{hybridTX}
\end{figure}

The transmitter block diagram of the proposed OFDM-HNIM system is displayed in Fig. \ref{hybridTX}. The $m$ incoming bits are partitioned to $G$ groups using bits splitter, each group contains $p=p_{1}+p_{2}+p_{3}$ bits, that are utilized to build an OFDM subblock of $L$ subcarriers length ($L=N_{F}/G$), where $N_{F}$ represents fast Fourier transform (FFT) size. The non-conventional bits ($p_{1}$ and $p_{2}$) represent the bits conveyed by the SNM and IM modulator. In each subblock, the SNM and IM mappers can be ordered arbitrary since the hybrid mapper combines both mappings without looking to the order of these non-convectional mappers. For convenient representation of the basic OFDM-HNIM system model, we represent the SNM bits by $p_1$, which are utilized by the SNM mapper to specify numerically the active OFDM subcarriers for each subblock, and $p_2$ bits, referred to IM bits, are exploited by the IM mapper to specify the indices of active subcarriers for each OFDM subblock. Then, the combined SNM and IM mapping in the hybrid mapper determines the subcarrier activation pattern (SAP) using a proper mapping technique.


For each subblock $g$ ($g=1,2,\ldots,G$), \textbf{index and number-dependent variable} $I \in[{1,2,\cdots,L}]$ which represents the set of activated subcarriers based on the hybrid mapper according to its corresponding $p_{1}$ and $p_{2}$ bits, and the remaining $L-I$ subcarriers are not active. It should be noted that $I$ is variable numeric representation for the active subcarriers according to $p_{1}$ and $p_{2}$ data code. The length of SNM {and IM} bits is set to be fixed {in order to simplify the system-level design.}
Table \ref{p1SAMEp2} presents a bits-to-SAP mapping for small values of $L=4$, $I \in[{1,2,3,4}]$ with $p_{1}=p_{2}=\log_{2}(L)= 2$ bits.
 The SAP for each subblock $g$ can be written as follows:
\begin{equation} \label{eq1cg} \mathbf{c}_g = \left[\begin{matrix} c_{1} & c_{2}\ \cdots \ c_{L} \end{matrix}\right]^{T}, \end{equation}
where $c_{i} \in {0,1}$ for $i = 1, 2, \cdots, L$. 
There {are} $p_{3} = I \log_{2}(M)$ bits {corresponding to specific conventional {constellation} symbols} {carried over $I$ active subcarriers}.

\begin{table*} [ht] 
\begin{center}
\caption{Look-up table of the proposed OFDM-HNIM scheme with $p_1=p_2=2$ bits}
\begin{tabular}{|c|c|c|c|c|}
\hline  $g$ & $p_1$ & $p_2$ & $\mathbf{c_g}$ & $I$  \\
\hline  1	& [0 0] & [0 0] & $[1\thinspace  0 \thinspace  0 \thinspace  0]^{T}$ & 1  \\
\hline  2	& [0 0] & [0 1] & $[0 \thinspace1 \thinspace0 \thinspace0]^{T}$ & 1  \\
\hline  3	& [0 0] & [1 0] & $[0 \thinspace0\thinspace 1\thinspace 0]^{T}$ & 1  \\
\hline  4	& [0 0] & [1 1] & $[0 \thinspace0\thinspace 0 \thinspace1]^{T}$ & 1  \\
\hline  5	& [0 1] & [0 0] & $[1 \thinspace1 \thinspace0 \thinspace0]^{T}$ & 2  \\
\hline  6	& [0 1] & [0 1] & $[1 \thinspace0\thinspace 1\thinspace 0]^{T}$ & 2  \\
\hline  7	& [0 1] & [1 0] & $[1\thinspace 0 \thinspace0 \thinspace1]^{T}$ & 2  \\
\hline  8	& [0 1] & [1 1] & $[0 \thinspace1 \thinspace0 \thinspace1]^{T}$ & 2  \\
\hline  9	& [1 0] & [0 0] & $[1 \thinspace1\thinspace 1\thinspace 0]^{T}$ & 3  \\
\hline  10	& [1 0] & [0 1] & $[1\thinspace 0 \thinspace1 \thinspace1]^{T}$ & 3  \\
\hline  11	& [1 0] & [1 0] & $[1 \thinspace1 \thinspace0 \thinspace1]^{T}$ & 3  \\
\hline  12	& [1 0] & [1 1] & $[0 \thinspace1\thinspace 1 \thinspace1]^{T}$ & 3  \\
\hline  13	& [1 1] & [0 0] & $[0 \thinspace0 \thinspace0 \thinspace0]^{T}$ & 0  \\
\hline  14	& [1 1] & [0 1] & $[0 \thinspace0 \thinspace1 \thinspace1]^{T}$ & 2  \\
\hline  15	& [1 1] & [1 0] & $[0 \thinspace1 \thinspace1 \thinspace0]^{T}$ & 2  \\
\hline  16	& [1 1] & [1 1] & $[1 \thinspace1\thinspace 1\thinspace 1]^{T}$ & 4  \\
\hline
\end{tabular} 
\label{p1SAMEp2}
\end{center}
\end{table*} 

It is shown from Table \ref{p1SAMEp2} that a subcarrier in a given index of $\mathbf{c_g}$ patterns is activated with an equal probability of 1/2. The presented codebook ($\mathbf{c_g}$) in Table \ref{p1SAMEp2} looks like a classical codebook generated in binary order; however, the proposed codebook ordering is different from the conventional binary ordering.
Moreover, the direct binary mapping does not provide flexibility, whereas, the proposed scheme can be seen as an adaptive and flexible transmission technique, whose mapping blocks can be employed 
based on the requirements of the used application. For example, individual SNM module could be used in the applications where higher throughput is needed at low modulation orders \cite{Jaradat2019}. Also, an IM module could be utilized for high EE applications \cite{Jaradat2019}. To have both high {energy} and spectral efficiencies in some applications, the combined SNM and IM modules could be exploited.      

The OFDM block in OFDM-HNIM system model is built as $\mathbf{x_F}=\begin{bmatrix} x_{F}(1) & x_{F}(2) &...& x_{F}(N_F) \end{bmatrix}$ based on the chain of $G$ subblocks. The remaining steps are done as in the classical OFDM system. 
By performing inverse fast Fourier transform (IFFT), the output vector is $\mathbf{x_t}$. With adding $N_{CP}$ cyclic prefix (CP) samples to the transmitted signal, the resultant would be $\mathbf{x_{CP}}=\begin{bmatrix} \mathbf{x_t} (N_{F}-N_{CP}+1:N_F) & \mathbf{x_t} \end{bmatrix}$. It is assumed that the transmitted signal passes to a fading channel with impulse response of $\mathbf{h_t}$, affected by additive white Gaussian noise (AWGN) with noise variance of $N_{o,T}$ in time domain. The received signal over the multi-path channel could be represented as:
\begin{eqnarray} \label{eq2yt} 
 \mathbf{y_t}=\mathbf{x_{t}} \circledast \mathbf{h_t}+\mathbf{n_t},
 \end{eqnarray}
where $\mathbf{n_t}\sim \mathit{\cal CN}(0,N_{o,T})$ represents the AWGN vector.\\ 
\subsection{OFDM-HNIM receiver}
The receiver part of the proposed OFDM-HNIM system is the reversal of the transmitter part, which includes CP removal, FFT processing, hybrid demapping, and detection. The resultant vector after taking off CP from the received signal can be written as $\mathbf{y_{NCP}}= \begin{bmatrix} y_{NCP}(0) & y_{NCP}(1) &...& y_{NCP}(N_F-1)\end{bmatrix}$.
Next, FFT operation is employed and the output signal is $\mathbf{y_F}$:

\begin{eqnarray} \label{eq3yF}
 \mathbf{y_F}=\mathbf{x_{F}} \mathbf{h_F}+\mathbf{n_F},
 \end{eqnarray}

Subsequently, a frequency domain equalizer of one tap is utilized to properly compensate the multi-path channel frequency selectivity as $\mathbf{y_{eq}}=\mathbf{y_F}/\mathbf{h_F}$, where $\mathbf{h_F}$ is the frequency response of the fading channel. Then, a maximum-likelihood (ML) detector is employed to extract the SAP. In order to regenerate $p_1$ and $p_2$, similar table is used at the receiver as the look-up Table \ref{p1SAMEp2} used for the transmitter. 
It should be noted that there would be similar SAP when $g=13,14, 15, 16$ as expected from their SNM mapping, due to setting a fixed length of $p_2$ data code. 
To solve this ambiguity in mapping, extra bits could be transmitted to receiver in order to differentiate between these exceptional cases. However, this solution is not spectrally efficient since extra non-data bits are transmitted which should be avoided. 
Moreover, the expected SNM mapping leads to a nonuniform subcarrier activation which results in unfair protection of transmitted bits, hindering the OFDM-SNM and OFDM-IM relevant schemes from attaining their ultimate error performance. 

The problem of unbalanced activation of subcarriers in the OFDM-SNM and OFDM-IM schemes is avoided in the proposed look-up Table {\ref{p1SAMEp2}} by having equiprobable probability in each subcarrier without dependency on the transmitter beforehand information about status of wireless channel \cite{WenESA2016}. Therefore, in order to avoid ambiguity and extra bits signaling at the ML detector as well as satisfying the equiprobable subcarrier activation (ESA) requirement \cite{WenESA2016}, these exceptional cases are assigned a unique SAP where their corresponding $I$'s are mostly different from what is expected from their SNM mapping. 

{However, the zero-active subcarrier dilemma, where all subcarriers are switched off and their corresponding data symbols can not be transmitted, can be seen in Table \ref{p1SAMEp2} at $g=13$ corresponding to $\mathbf{c_g}=[0\thinspace  0\thinspace  0\thinspace  0]^T$. This dilemma could lead to difficulty in higher layer design, especially the synchronization process which is the key for OFDM-based systems in order to reduce ICI and/or inter symbol interference (ISI) \cite{Basar_trans_2013,8241721}. One efficient solution to such a dilemma could be done by exploiting the lexicographic ordering principle and implementing the codebook optimization method as in \cite{8519769}, where the codebooks are optimized in such a way that the subcarriers are activated based on their corresponding instantaneous CSI. The novel codebook design in \cite{8519769} does not involve optimizing the power usage unlike the forward error correction method that adds more computational complexity at the receiver \cite{EnhancedSIMOFDM2011}. 

As shown in Table \ref{p1SAMEp2}, the equal probable bit sequence enables easier detection and designs of higher layer protocols unlike the dual-mode transmission protocol \cite{8241721}, where the modulation has been employed on a bit sequence with variable length. Control signaling is not required in \cite{8519769} unlike the method in \cite{DFOFDMIM2018} where always-active control subcarrier has been used for control signaling transmission. Therefore, the lexicographic-based codebook design in \cite{8519769} provides efficient solution to the observed dilemma as compared to the existing solutions \cite{EnhancedSIMOFDM2011,8241721,DFOFDMIM2018}. In this paper, we focus on the basic idea and system model of the proposed OFDM-HNIM, whereas employing the novel design in \cite{8519769} to the proposed scheme could be considered as a future work in order to improve the system performance.}\footnote{In this paper, our focus is introducing a new physical layer transmission scheme. The usage of the proposed modulation scheme at the upper layers is out of this paper scope.}

{The remaining steps of the OFDM-HNIM receiver are:} The bits conveyed by index and number of subcarriers in each subblock are estimated based on the obtained SAP using hybrid demapper which represents the contrary of the hybrid mapper used at the transmitter. Then, constellation symbols detection is performed based on the received SAP for each subblock. Finally, the detected bits from the hybrid demapping and conventional QAM detection are jointly formed the latest estimated subblock bits. By performing similar operations to all subblocks, the recovered data sequence is acquired for the whole OFDM block.
\section{Performance evaluation of the proposed hybrid scheme} 
Here, evaluation of the hybrid system performance is performed based on some metrics such as spectral efficiency, average bit error probability, {energy} efficiency, and computational complexity.   


\subsection{Spectral Efficiency}
The achievable rate of the proposed OFDM-HNIM scheme can be calculated based on the mutual information between the channel input and the channel output averaged over the subcarriers \cite{WenRate2016}. Since the transmission in the proposed OFDM-HNIM scheme is performed in a subblock level and we treat the subblocks independently due to having different subblocks realizations. Therefore, the mutual information over the OFDM-HNIM block should be calculated in a subblock level as shown in (\ref{rateOFDMHNIM}) which presents the achievable rate of the proposed OFDM-HNIM scheme:

\begin{eqnarray} \label{rateOFDMHNIM}
\begin{split}
R_{h}=\frac{\mathbf{I(\mathbf{x_{F}^g})}}{L}=\frac{\mathbf{H(\mathbf{x_{F}^g})}}{L}+\frac{\mathbf{H(\mathbf{x_{F}^g}|\mathbf{y_{F}^g})}}{L},
\end{split}
\end{eqnarray}
where $\mathbf{I(\mathbf{x_{F}^g})}$ represents the mutual information of the channel input of the $g$-th subblock ($\mathbf{x_{F}^g}$), $\mathbf{H(\mathbf{x_{F}^g})}$ is the marginal entropy of $\mathbf{x_{F}^g}$ and $\mathbf{H(\mathbf{x_{F}^g}|\mathbf{y_{F}^g})}$ is the conditional entropy of $\mathbf{x_{F}^g}$ and the channel output of the $g$-th subblock ($\mathbf{y_{F}^g}$).

The achievable rate formula in (\ref{rateOFDMHNIM}) can be written in terms of probability density functions (PDFs) of the channel input, channel, and channel output as follows \cite{EIM2017}:

\begin{equation} \label{pdfOFDMHNIM}
\begin{split}
R_{h}=\eta_{h}-\frac{1}{2^{p} L} \sum_{j=1}^{2^{p}}  E_{\mathbf{h_{F}}^g}&\Bigg\{
\int f(\mathbf{y_{F}}^g|\mathbf{x_{F}}^{g(j)},\mathbf{h_{F}}^g)\\ & \times \log_{2}\Bigg({\dfrac{f(\mathbf{x_{F}}^{g(j)},\mathbf{y_{F}}^g|\mathbf{h_{F}}^g)}{f(\mathbf{y_{F}}^g|\mathbf{h_{F}}^g)}}\Bigg) d\mathbf{y_{F}}^g\Bigg\},
\end{split}
\end{equation} 
where $f(\mathbf{y_{F}}^g|\mathbf{h_{F}}^g)=\dfrac{1}{2^p} \sum_{j=1}^{2^{p}} f(\mathbf{y_{F}}^g|\mathbf{x_{F}}^{g(j)},\mathbf{h_{F}}^g)$.

Furthermore, (\ref{pdfOFDMHNIM}) can be simplified to the following in the analogy with \cite{WenRate2016,EIM2017}:

\begin{eqnarray}
R_{h}\approx \eta_{h}-\frac{1}{2^{p} L} \sum_{j=1}^{2^{p}} \log_2{\sum_{w=1}^{2^{p}} \frac{1}{det(\mathbf{I_{L}}+\mathbf{K_{L}} \mathbf{U}_{j,w})}},
\end{eqnarray}
where $\eta_{h}$ is the SE of the proposed hybrid scheme, and it can be found in (\ref{eq4SE}), $\mathbf{K_L}=E[\mathbf{h_{F}^{g}} \mathbf{(h_{F}^{g})^H}]$ is the covariance matrix of $\mathbf{h_{F}^{g}}$, $\mathbf{I_{L}}$ represents the identity matrix with dimensions $L \times L$, and $\mathbf{U}_{j,w}=\dfrac{(\mathbf{x}_{F}^{g(j)}-\mathbf{x}_{F}^{g(w)})^H (\mathbf{x}_{F}^{g(j)}-\mathbf{x}_{F}^{g(w)})}{2 N_{o,F}}$, where $\mathbf{x}_{F}^{g(j)}$ and $\mathbf{x}_{F}^{g(j)}$ represent the $j$-th and $w$-th realization of the subblock $\mathbf{x}_{F}^{g}$, respectively. It should be noted that $N_{o,F}$ is the noise variance in frequency domain. This analytical result will be verified by simulation as well.

The maximum achievable rate (i.e. SE) of the proposed OFDM-HNIM scheme is:
\begin{eqnarray}  \label{eq4SE}
 {\eta_{h}}=\dfrac{\sum_{g=1}^{G}\left( \log_{2}(L)+\log_{2}{L\choose I(g)}+I(g) \log_2(M)\right) }{N_{F}+N_{CP}}, 
 \end{eqnarray}
where $I(g)$ represents a numeric variable of SAP in each subblock of size $L$. Since two parts of an incoming non-conventional bits are used for hybrid mapping compared to only one part of OFDM-SNM and OFDM-IM schemes. Thus, the hybrid scheme improves the SE over the OFDM-IM and OFDM-SNM transmission schemes. For example, Table \ref{SEcompOFDM} shows the maximum achievable rate or SE for the proposed hybrid scheme and its competitive schemes including OFDM-SNM, OFDM-IM, and conventional OFDM, assuming $N_F=64$ with $N_{CP}=8$, and BPSK modulation is employed. It should be noted that fixed length of $p_2$ bits in the hybrid mapping is considered for convenient comparison. Also, the number of subblocks ($G$) equals to either 16 when $L=4$ ($G=N_F/L=64/4=16$), or ($G=N_F/L=64/8=8$) when $L=8$. {For the system settings shown in Table \ref{SEcompOFDM}}, the proposed hybrid scheme outperforms its counterparts in terms of SE . {Moreover,} the SE becomes much more significant in the hybrid scheme as the subblock length increases. This SE trend can not be achieved by the conventional OFDM-SNM and OFDM-IM schemes in which SE is degraded as subblock size rises.

\begin{figure}[t]
\centering
\includegraphics[height=2.3in,width=3in]{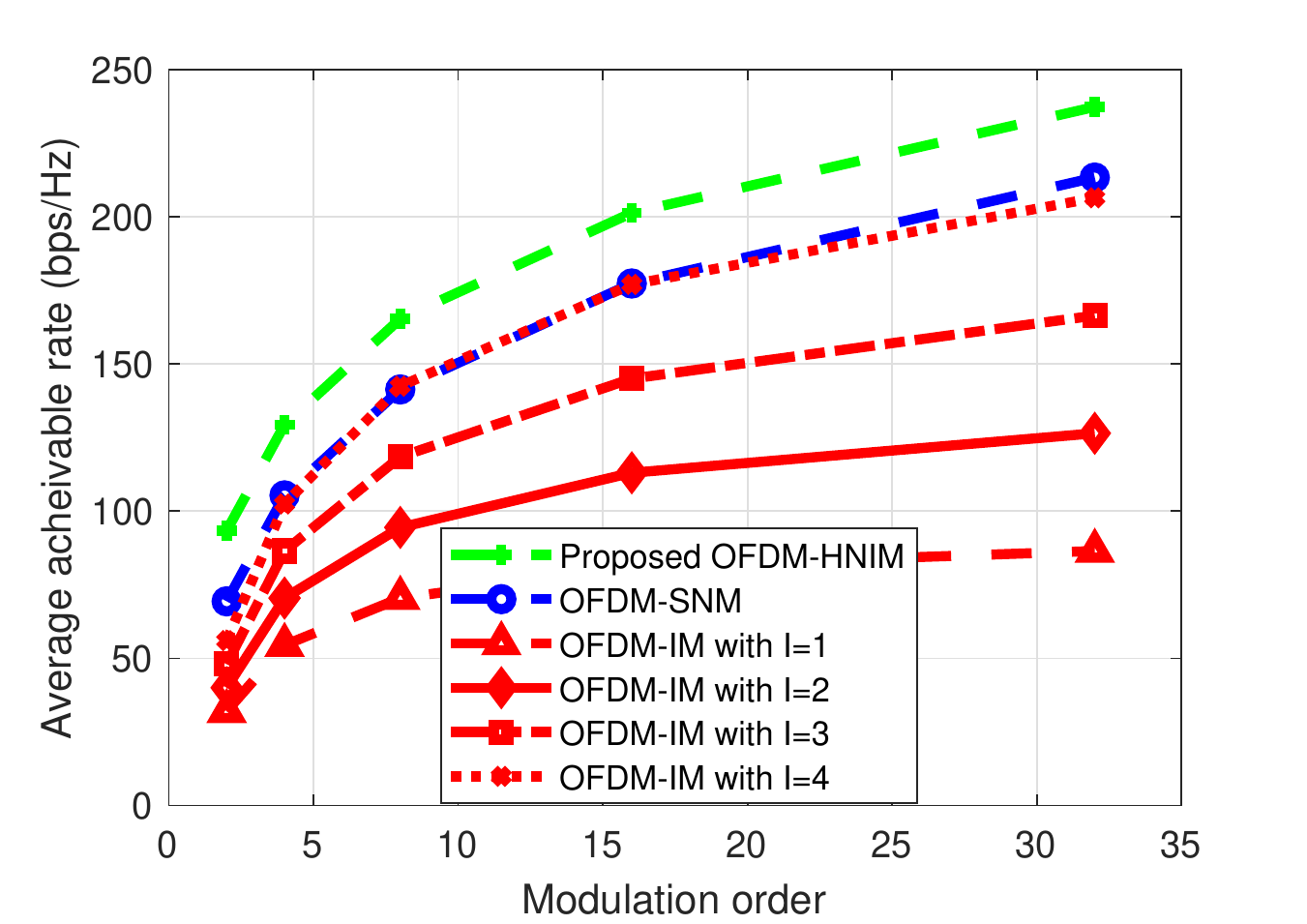}

\caption{Average achievable rate of the featured OFDM-based modulation options as function of $M$ when the subblock size set to 8.}
\label{avgSEcompdiffM}
\end{figure}

Table \ref{SEcompOFDM} shows the SE gain of different OFDM-based modulation schemes under BPSK modulation {relative to classical OFDM system as a baseline}\footnote{The phrase N/A, as shown in Table \ref{SEcompOFDM}, refers to not applicable.}. It is clear from Table \ref{SEcompOFDM} that the proposed OFDM-HNIM outperformed its counterparts in terms of SE at low modulation order ($M=2$). A comprehensive comparison in terms of the average achievable rate between the featured OFDM-based modulation options when the subblock size of 8 is shown in Fig. \ref{avgSEcompdiffM} where hybrid scheme outperforms its counterparts including conventional OFDM-SNM and OFDM-IM schemes at different modulation orders.

Mathematically, we can deduce the values of $M$ and $L$ that satisfy the condition where the average achievable rate of the proposed scheme improves over that of the conventional OFDM. The average achievable rate of the hybrid scheme can be calculated from (\ref{eq4SE}) with average number of active subcarriers of $I_{avg}=(L+1)/2$ over the available OFDM subcarriers:

\begin{eqnarray} \label{SEHYBRID}
 {\bar{\eta}_{h}}=\dfrac{N_{F} \left( \log_{2}(L)+\log_{2}{L\choose I_{avg}}+I_{avg} \log_2(M)\right)}{L (N_{F}+N_{CP})}.
\end{eqnarray}
By comparing (\ref{SEHYBRID}) to the average achievable rate of the conventional OFDM which is: 

\begin{eqnarray} \label{SEOFDM}
 {\bar{\eta}_{OFDM}}=\dfrac{N_{F}\log_{2}(M)}{N_{F}+N_{CP}}, 
\end{eqnarray}
we observe the trend of $M$ and $L$ for the hybrid scheme and conventional OFDM, then we can set the inequality results from considering the average achievable rate of the hybrid scheme greater than that of the conventional OFDM as follows: 

\begin{eqnarray} \label{SEinequlaity}
 \log_{2}(L)+\log_{2}{L\choose I_{avg}}+I_{avg} \log_2(M) \geq L \log_2(M). 
\end{eqnarray}

\begin{figure}[t]
\centering
\includegraphics[height=2.3in,width=3in]{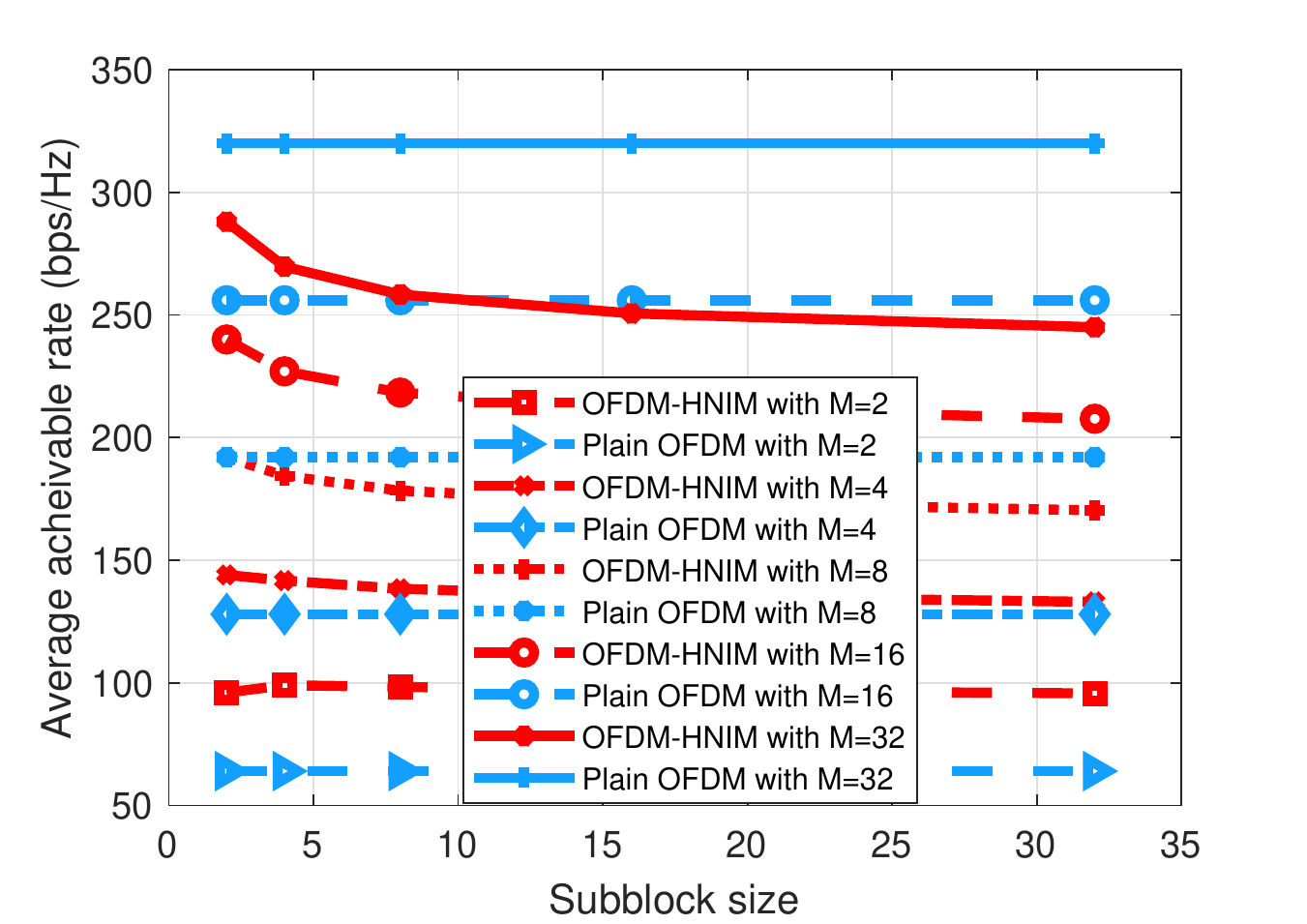}

\caption{Average achievable rate comparison between the hybrid scheme and plain OFDM in terms of subblock size.}
\label{avgSEdiffL}
\end{figure}

Fig. \ref{avgSEdiffL} shows that the average achievable rate of the conventional OFDM does not depend on the subblock size since subblock-based transmission is not employed by the plain OFDM. However, the hybrid scheme outperforms the plain OFDM at low modulation orders, specifically when $M=2$ or $M=4$, for different subblock sizes. However, as $M$ increases, it is less likely that the proposed scheme has a higher average achievable rate than the conventional OFDM, and this rate degradation also increases as subblock size increases. This happens because of having more probable bits experience deep fading as subblock size increases and it becomes difficult for receiver to detect these bits in deep-fading condition.

\begin{table*}[]
\begin{center}
\caption{SE of the featured OFDM-based modulation schemes with BPSK} 
\begin{tabular}{|l|l|l|l|l|l|l|}
\hline
OFDM-based Modulation Scheme & $L$ & \textbf{$G\times p_{1}$} & \textbf{$G\times p_{2}$} & \textbf{$G\times p_{3}$} & \textbf{$SE=G\times p$} & SE gain over conventional OFDM \\ \hline
\multirow{2}{*}{Proposed OFDM-HNIM scheme} & 4 & 32 & 32 & 32 & 96 & 1.333 \\ \cline{2-7} 
 & 8 & 24 & 24 & 216 & 264 & 3.6667 \\ \hline
\multirow{2}{*}{OFDM-SNM \cite{Jaradat2018}} & 4 & 32 & 40 & N/A & 72 & 1 \\ \cline{2-7} 
 & 8 & 24 & 36 & N/A & 60 & 0.8333 \\ \hline
\multirow{2}{*}{OFDM-IM \cite{Basar_trans_2013}} & 4 & 32 & 32 & N/A & 64 & 0.8889 \\ \cline{2-7} 
 & 8 & 24 & 24 & N/A & 48 & 0.6667 \\ \hline
\multirow{2}{*}{Conventional OFDM} & 4 & N/A & N/A & N/A & 72 & 1 \\ \cline{2-7} 
 & 8 & N/A & N/A & N/A & 72 & 1 \\ \hline 
\end{tabular}
\label{SEcompOFDM}
\end{center}
\end{table*}

\subsection{Average Bit Error Probability} 
The whole block in the hybrid scheme should be detected based on transmission bits conveyed in the number and index of activated subcarriers and the conventional constellation symbols \cite{8241721}. 
So, the fundamental block error rate (BLER) is computed first and then the average BER calculated based on the average BLER \cite{8519769,8614439}. Actually, BLER is basically expressed in terms of pairwise error probability (PEP). The PEP is calculated in two steps, first with the effect of channel ($\mathbf{h_t}$) in which PEP in this case is called conditional PEP, and the second step is done by getting rid of channel to obtain the unconditional PEP.  
The conditional PEP could be represented by using the Q-function \cite{Jafarkhani2005}:
\begin{equation} \label{ABEPCOND}
P(\mathbf{c}_g\longrightarrow \mathbf{\hat{c}}_{\hat{g}}|\mathbf{h_t}) =Q( \sqrt{ \frac{P_t}{N_{o,F}}{||\mathbf {h_t} (\dfrac{\mathbf{c}_g}{\sqrt{I(g)}}-\dfrac{\mathbf{\hat{c}}_{\hat{g}}}{\sqrt{I(\hat{g})}})||^{2}} } ),\\   
\end{equation} 
where $P_t$ represents the total transmit power, $I(g)$ and $I(\hat{g})$ represent the number of active subcarriers in the $g$-th and $\hat{g}$-th subblock, respectively, $\mathbf{c}_g$ and $\mathbf{\hat{c}}_{\hat{g}}$ are transmitted and detected sequences, respectively.
The conditional PEP shown above can be approximated by using the exponential approximation of the Q-function as \cite{Chiani2002,1210748}:
\begin{equation} \label{QFUNCTION}
Q(x)\approx \sum_{j=1}^{2}\rho_j \exp(-\eta_j x^2),  
\end{equation} 
where $\rho_1=1/12$ and $\rho_2=1/4$, $\eta_1=1/2$ and $\eta_2=2/3$. The new formula of conditional PEP becomes \cite{8703169}:

\begin{equation} \label{SUMPROD}
P(\mathbf{c}_g\longrightarrow \mathbf{\hat{c}}_{\hat{g}}|\mathbf{h_t}) =\sum_{j=1}^{2}\rho_j \prod_{l=1}^L \exp(\frac{-\eta_j P_t}{N_{o,F}} R(l) \Delta(l,g,\hat{g})),
\end{equation} 
where $R(l)=|h_t(l)|^2$ and $\Delta(l,g,\hat{g})=|\dfrac{\mathbf{c}_g(l)}{\sqrt{I(g)}}-\dfrac{\mathbf{\hat{c}}_{\hat{g}}(l)}{\sqrt{I(\hat{g})}}|^2$.

The unconditional PEP can be found by averaging the conditional PEP over the channel as \cite{7544555,8006223}:
\begin{equation} \label{ABEPUNCOND}
P(\mathbf{c}_g\longrightarrow \mathbf{\hat{c}}_{\hat{g}}) =\sum_{\mathbf{\hat{c}}_{\hat{g}} \neq \mathbf{c}_g}E_\mathbf{h_t}\big\{P(\mathbf{c}_g\longrightarrow \mathbf{\hat{c}}_{\hat{g}})|\mathbf{h_t})\big\},    
\end{equation} 





With the assistance of the aforementioned PEP and assuming equiprobable information bits, the upper bound of ABEP in the hybrid system could be obtained \cite{6877687,Basar_trans_2013}:
\begin{eqnarray} \label{TWOSUM}
P_{b}(E)=\frac{1}{p_{g}\;n_{x}} \sum_{g=1}^{G}\sum_{\mathbf{\hat{c}}_{\hat{g}}\neq\mathbf{c}_g}P(\mathbf{c}_g\longrightarrow\mathbf{\hat{c}}_{\hat{g}}) e(\mathbf{c}_g,\mathbf{\hat{c}}_{\hat{g}}),
   \end{eqnarray}
 where $p_{g}$ represents the length of vector that contains information bits corresponding to OFDM subblock, $n_{x}$ represents numerically the possible realizations of the transmitted sequence, and $e(\mathbf{c}_g,\mathbf{\hat{c}}_{\hat{g}})$ denotes errors in information bits due to erroneously choosing $\mathbf{\hat{c}}_{\hat{g}}$ rather than $\mathbf{c}_g$ \cite{Basar_trans_2013}. This theoretical BER result will be verified by simulation as well. 

 






\subsection{Energy Efficiency} 
{ The energy efficiency (EE) implies how efficiently energy is consumed in a given system. 
The EE of the proposed OFDM-HNIM scheme is analyzed in terms of the energy saving factor ($ESF$) achieved by not activating all of the available subcarriers in the OFDM block. The evaluation of $ESF$ for the featured schemes is discussed here.}
 
 { One main aspect of the overall energy saving is the saving in signal transmission \cite{8663599}. The EE in terms of $ESF$ depends on the ratio of saved energy {under} a given total transmit power ($P_t$) in {a} single OFDM symbol. We assume equiprobable subcarrier
activation \cite{WenESA2016} for convenient comparison between the proposed OFDM-HNIM scheme and its competitive schemes. Moreover, the average number of active subcarriers is considered {in our} {EE} analysis. The number of activated subcarriers in an OFDM block impacts both SE and EE, and their ratios relation with respect to the traditional OFDM as a reference scheme \cite{OFDMimprovedGIM2016}, {can} be defined as \cite{8663599}}
\begin{equation} \label{EEeq}
    EE_{r}=\frac{SE_{r}}{1-ESF},          
\end{equation}
{where $EE_{r}$ and $SE_{r}$ represent the EE and SE ratios of the scheme of interest with respect to the conventional OFDM scheme. Maximizing SE and EE are generally two conflicting objectives as can be clearly seen from (\ref{EEeq}) \cite{6108331}.
$ESF=N_{a}/N_{v}$ represents the saved energy by activating $N_{a}$ subcarriers out of $N_{v}$ available subcarriers.}

\begin{figure}[t]
\centering
\includegraphics[height=2.3in,width=3in]{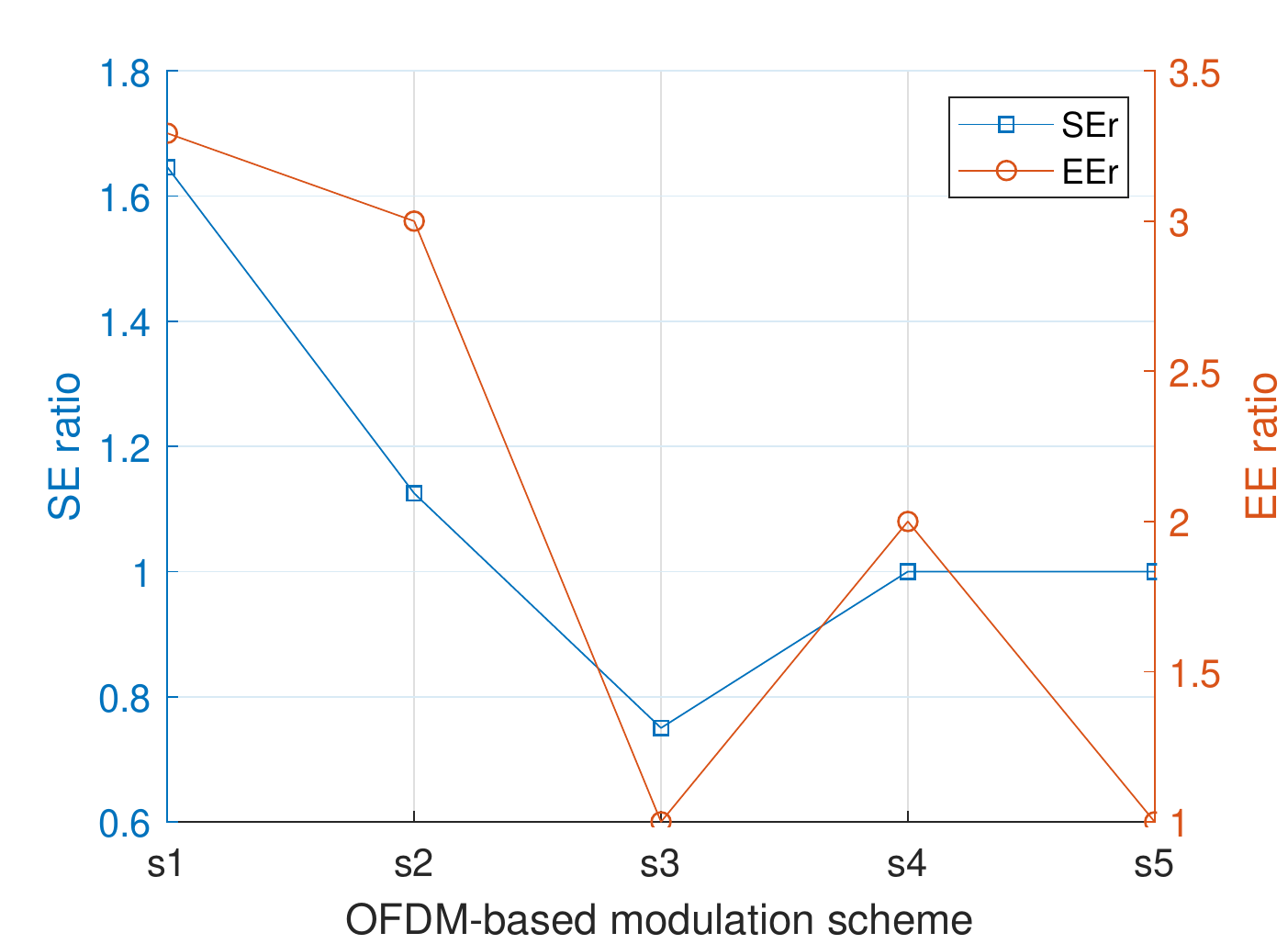}

\caption{The SE and EE ratios of the featured OFDM-based modulation options with respect to the plain OFDM. {The symbols s1, s2, s3, s4, s5 correspond to the proposed OFDM-HNIM scheme, OFDM-SNM, OFDM-IM with $AR = 0.25$, OFDM-IM with $AR = 0.5$, and conventional OFDM, respectively.}}
\label{SErEEr}
\end{figure}

{For example, in the proposed OFDM-HNIM scheme, half of subcarriers are active in average which is the equivalent to the case of OFDM-IM scheme with half of its available subcarriers are active, i.e, subcarrier activation ratio of half ($AR=0.5$). The energy saving resulting from the OFDM-HNIM and OFDM-IM with $AR=0.5$ is limited to half of the OFDM symbol ($ESF=0.5$). 
In other words, the active subcarriers in OFDM-HNIM scheme gain higher power as compared to the individual subcarriers in classical OFDM under the same transmitted power.
Hence, the receiver design of the proposed OFDM-HNIM scheme would only differentiate between the active group with higher power and the inactive group regardless of the exact amplitude and phase of each subcarrier.

On the other hand, the $ESF$ trend in the OFDM-SNM and OFDM-IM depends on the subblock length and $AR$. It is well-known that the conventional OFDM is not {an} EE scheme since most of the available subcarriers are usually used for data transmission which results in very low $ESF$.} 
{Fig. \ref{SErEEr} shows the $SE_r$ and $EE_r$ of the featured OFDM modulation schemes with respect to classical OFDM. 
It is clear from Fig. \ref{SErEEr} that the proposed OFDM-HNIM scheme provides higher $SE_r$ and $EE_r$ as compared to the OFDM-SNM, OFDM-IM with low $AR$, and the classical OFDM.}

\subsection{Complexity Analysis}
In the proposed hybrid scheme, the detection for the number and indices of active subcarriers as well as the conventional QAM symbols can be performed subblock by subblock without introducing any performance loss since the encoding processes for all subblocks are independent. Based on ML, a joint decision on the set of active subcarriers as well as the conventional QAM symbols in each $g$-th subblock as:

\begin{eqnarray}  \label{optimalMLeq}
\min\limits_{s_{g}, I_{g}} ||\mathbf{y_{F}}- \mathbf{x_{F}} \mathbf{h_{F}}||^2,
   \end{eqnarray}
where $s_{g}$ represents the conventional QAM symbol carried over the active subcarrier in the $g$-th subblock. The computational complexity of this optimal ML detector in terms of complex multiplications is of order $\sim \mathcal{O}(G \thinspace M^{n/2})$ per bit detected in each OFDM-HNIM block, which becomes impractical for large values of $G$, $n$ and $M$, due to its exponentially increasing complexity.


To simplify the bulky search in this optimal ML detection, two different ML detectors have been implemented; Perfect subcarrier activation pattern estimation (PSAPE)-based, and imperfect SAP estimation (ISAPE)-based ML detectors. The PSAPE-based detector ignores the errors due to the incorrect detection of subcarriers in the received SAP. 
However, in the ISAPE-based detector, the conventional QAM demodulation phase would be unsuccessful to deduce $p_3$ bits correctly for the erroneous SAP detection case. This happens due to demodulation of some unmodulated subcarriers. Therefore, the transmitted $p_1$, $p_2$, and $p_3$ bits are erroneous when the detected SAP is wrong. On the other hand, if the SAP is correctly detected, then, $p_1$ and $p_2$ is correct but it is not certain that $p_3$ is correctly estimated. Actually, in case of erroneous conventional symbol detection for the received active subcarriers, $p_3$ bits are affected. Otherwise, a correctly symbol detection for activated subcarriers in a correctly estimated SAP produces a correct estimation of $p_1$, $p_2$, and $p_3$ bits. The PSAPE-based detector is considered in the following simulations due to its low-complexity compared to that of ISAPE-based detector. It should be noted that the performance comparison between both detectors is shown as well. 
{On the other hand, LLR detector is also employed with the proposed OFDM-HNIM scheme in order to reduce the computational complexity at the receiver.}

To compare the computational complexity of the these proposed ML-based detectors, namely PSAPE and ISAPE-based detectors, with that of the optimal ML detector {and LLR detector,} we consider the average number of metric calculations per subcarrier as a performance metric. Table \ref{complexitycomp} presents the complexity comparison results for the featured OFDM-based modulation schemes. As seen from Table \ref{complexitycomp}, the computational complexity of the optimal ML detector is highly susceptible to parameters $G$, $n$, and $M$; however, the complexity of the two proposed ML-based detectors is only determined by $G$ and, apparently, much lower than the optimal ML-based detector, {and there is much reduction in complexity by employing LLR detector in order of $M$}. Moreover, Table \ref{complexitycomp} shows the comparison between the detection complexity of the proposed hybrid scheme with that of its counterparts schemes such as OFDM-SNM, OFDM-IM, and plain OFDM. {It is clear that the proposed OFDM-HNIM along with LLR detector is considered as a low-complex scheme with similar order as that in OFDM-IM and conventional OFDM.}

\begin{table*}[]
\begin{center}
\caption{Complexity comparison between different detectors for the featured OFDM-based modulation schemes}
\begin{tabular}{|l|l|l|}
\hline
\begin{tabular}[c]{@{}l@{}}{OFDM-based} Modulation Scheme\end{tabular} & Detector type & Complexity order \\ \hline
\multirow{4}{*}{Proposed OFDM-HNIM scheme} & Optimal ML & $\sim \mathcal{O}(G \thinspace M^{n/2})$ \\ \cline{2-3} 
 & PSAPE & $\sim \mathcal{O}(G)$ \\ \cline{2-3} 
 & ISAPE & $\sim \mathcal{O}(G)$ \\ \cline{2-3} 
 & LLR & $\sim \mathcal{O}(M)$ \\ \hline
OFDM-SNM & ML & $\sim O(L,I,M)$ \\ \hline
\multirow{2}{*}{OFDM-IM} & \multirow{2}{*}{Near optimal LLR} & \multirow{2}{*}{$\sim O(M)$} \\
 &  &  \\ \hline
\multirow{2}{*}{Conventional OFDM} & \multirow{2}{*}{ML} & \multirow{2}{*}{$\sim O(M)$} \\
 &  &   \\ \hline
\end{tabular}
\label{complexitycomp}
\end{center}
\end{table*}

\section{Simulation Results} \label{simulationSection}
Here, the throughput, BER, and {EE} performances of the hybrid system are compared to those in OFDM-GIM with $I=[1,2,3]$ and $L=4$ \cite{Fan2015}, OFDM-IM, OFDM-SNM, and conventional OFDM. It is assumed that FFT size ($N_F$) is 64, subblock length ($L$) is 4, so the number of subblocks $G=N_F/L=16$, and SNM and IM bits are $p_1=p_2=\log_2(L)=2$ bits in each subblock. The modulation used in the simulation is BPSK.
The employed frequency-selective channel is Rayleigh distributed with 10 taps and similar SNR (or $E_b/N_{o,T}$) is considered in order to have fair comparison between the considered schemes, where $E_b$ is the bit energy. It is assumed that the CP length ($N_{CP}$) is longer than {the effective channel impulse response}. Moreover, we assume that the {CSI} is not available at the transmitter.


\begin{figure}[t]
\centering
\includegraphics[height=2.3in,width=3.2in]{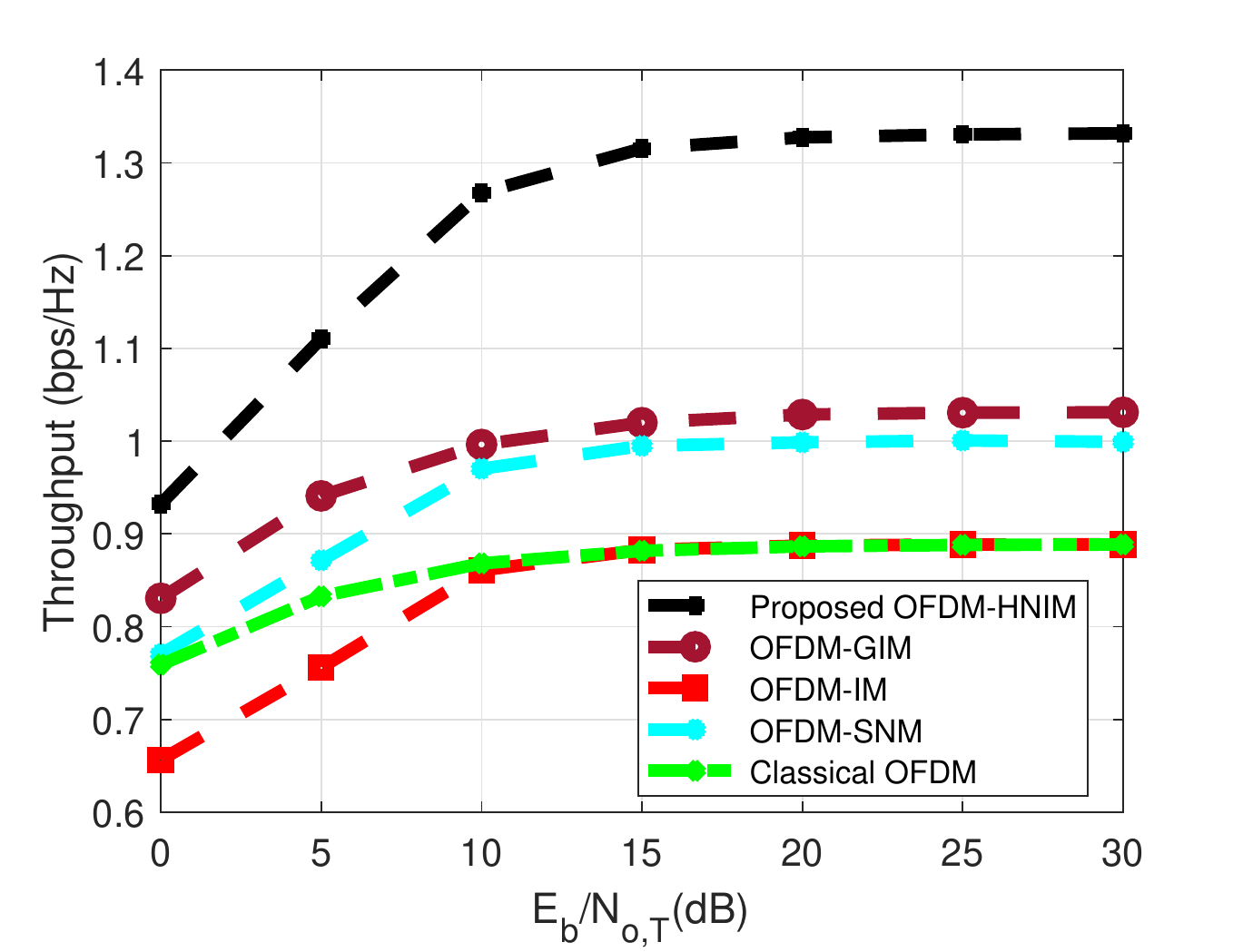}

\caption{Throughput of the proposed scheme and its competitive schemes under BPSK.}
\label{THRBPSK}
\end{figure} 

\begin{figure}[t]
\centering

\includegraphics[height=2.3in,width=3.2in]{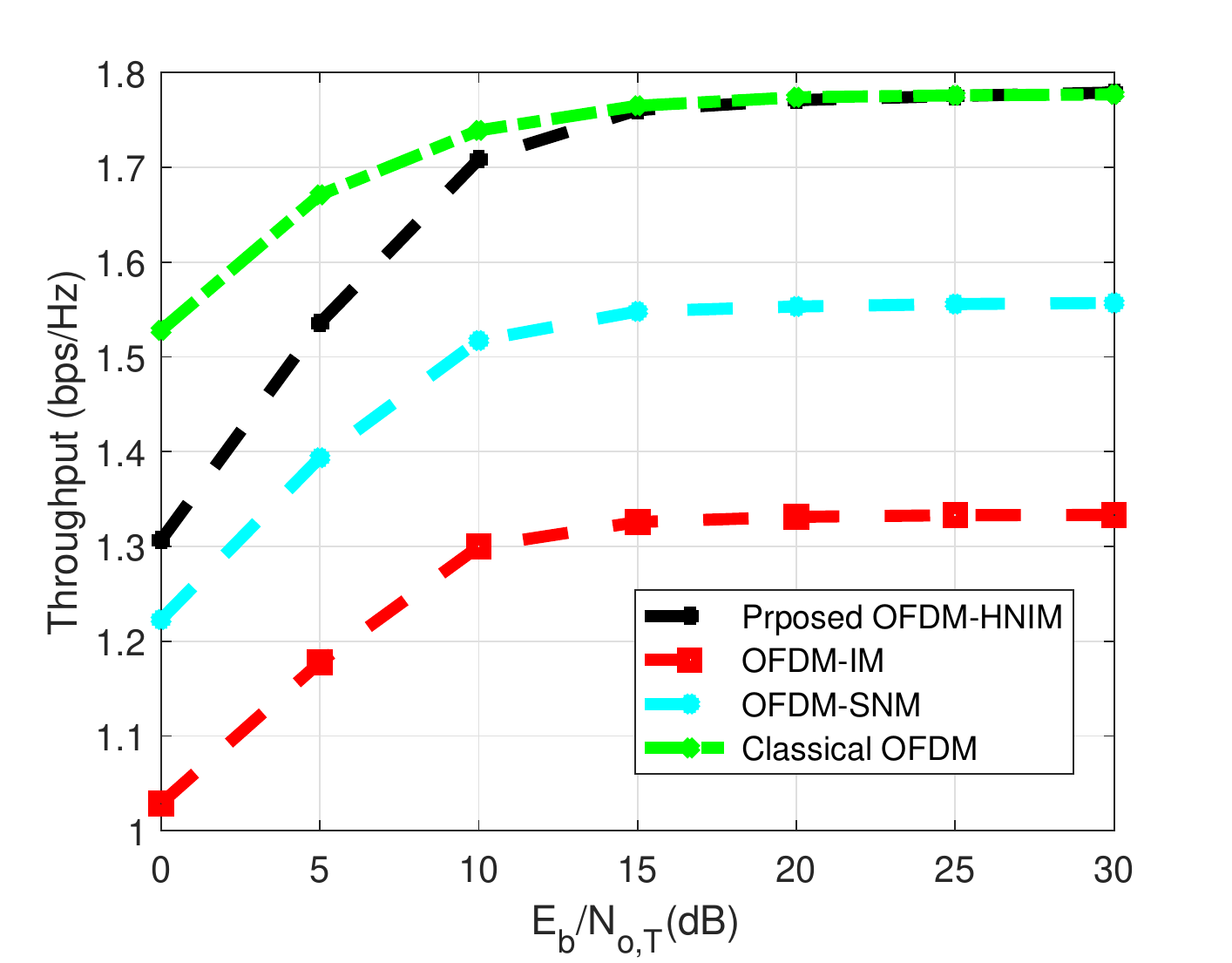}

\caption{Throughput of the proposed scheme and its competitive schemes under QPSK.}
\label{THRQPSK}
\end{figure}

The throughput performances of the proposed hybrid scheme compared to its competitive schemes under different modulation types are shown in Fig. \ref{THRBPSK} and Fig. \ref{THRQPSK}. 
Here, the simulated throughput metric is found by multiplying the SE for a given modulation scheme by simulated BER subtracted from one.
The saturation points of the achievable rates of all the schemes or their SEs fall at very high SNR. These SE values dependent on the mapping between incoming bits, SAPs and conventional symbols, as shown in (\ref{eq4SE}).

At low order modulation especially BPSK, as shown in Fig. \ref{THRBPSK}, the SE of the hybrid scheme is 1.33 bits/s/Hz with SE gain of 0.44 compared to OFDM-IM, and conventional OFDM, 0.33 compared to OFDM-SNM, and 0.3 compared to its OFDM-GIM counterpart. The reason behind the superiority of the hybrid scheme over its competitive schemes under BPSK is because of low noise and interference effects in lower order modulation as well as additional information sent by index and number of active subcarriers. This obviously shows that the hybrid scheme is an improved spectrally efficient scheme compared to its OFDM-IM and OFDM-SNM counterparts where only the indices or numbers of active subcarriers are exploited to convey additional data bits.

{The throughput performances for the featured OFDM-based schemes under QPSK is shown in Fig. \ref{THRQPSK} where SE superiority of the OFDM-HNIM scheme is also achieved over its conventional OFDM-SNM and OFDM-IM schemes. However, the classical OFDM outperforms the OFDM-HNIM in terms of throughput under QPSK especially at low SNR values due to sparse distribution of active subcarriers featured in the proposed hybrid scheme, classical OFDM-SNM and OFDM-IM schemes.}

\begin{figure}[t]
\centering
\includegraphics[width=3.4in]{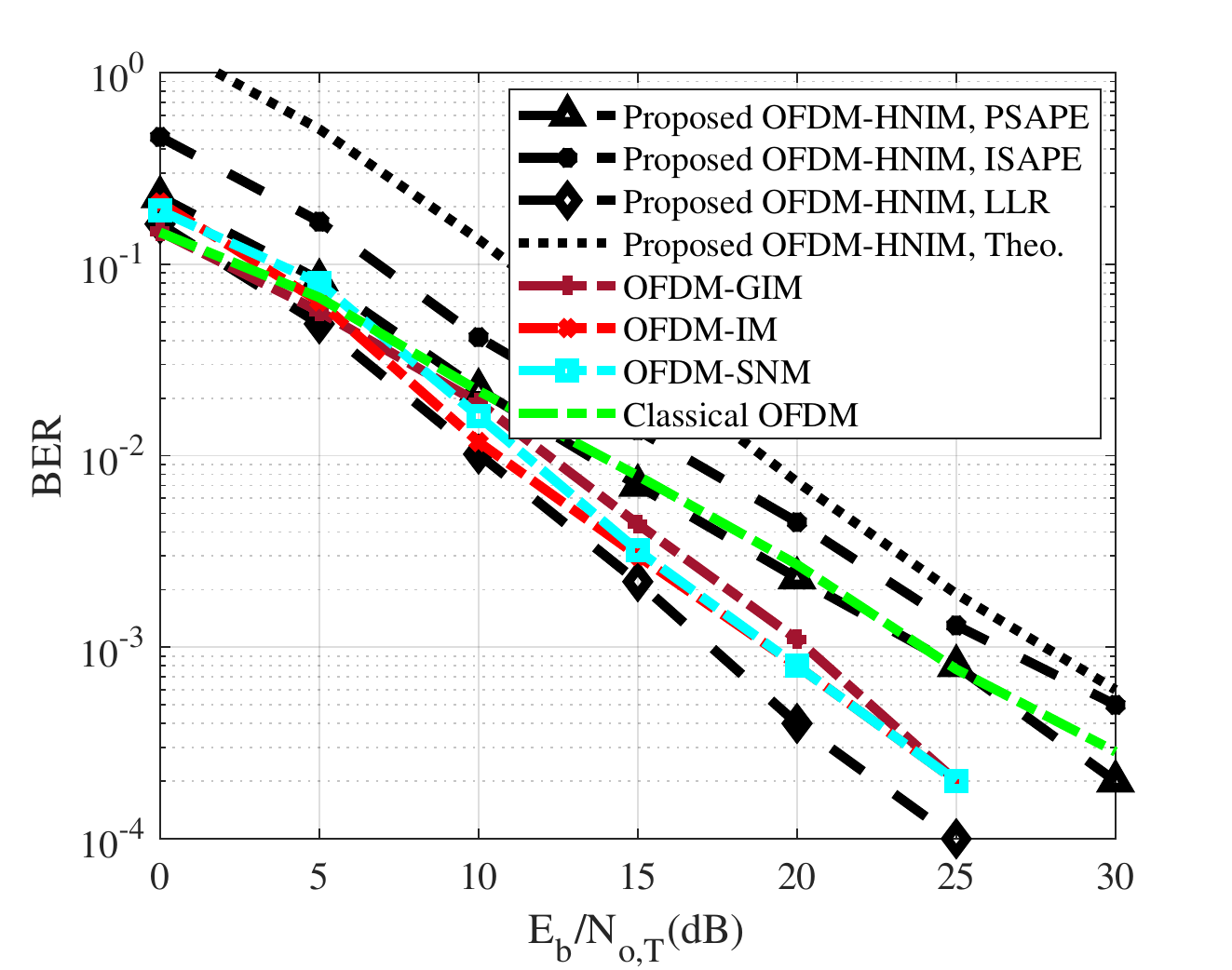} 

\caption{BER of the proposed scheme, OFDM-GIM, OFDM-IM, OFDM-SNM, and classical OFDM under frequency-selective Rayleigh channel with BPSK.}
\label{BERALL}
\end{figure}

The BER performance {for the featured OFDM-based modulation options} when employing a certain conventional modulation type such as BPSK, is shown in Fig. \ref{BERALL}. It is assumed that the same total power is allocated at the transmitter of each of the considered schemes, and the original power of the inactive subcarriers is evenly reallocated for the active ones.
Due to the three different sources of errors occurred by using the ISAPE-based detector in the hybrid system, its BER performance would be worse than that of PSAPE detector. 

As can be seen from Fig. \ref{BERALL} that {the LLR detector of the proposed OFDM-HNIM achieves improved BER performance compared to its competitive schemes.}
{However, employing the two proposed ML-based detectors results in} less than 1 dB coding loss in the hybrid scheme compared to its competitive schemes at low SNR values, and more SNR loss in high SNR region. The reason behind this is the weaker protection featured in the number-modulated and index-modulated bits compared to the conventional modulation bits at higher values of SNR, therefore the number-modulated and index-modulated bits are more probable to encounter significant error transmission in the high SNR region. Moreover, ratio of number-modulated and index-modulated bits in the proposed hybrid scheme is larger than in its competitive schemes, that increases the influence of active number and index bits on the BER more.
In Fig. \ref{BERALL}, the derived theoretical ABER of the proposed scheme becomes considerably tight with computer simulations as SNR increases. 

\begin{figure}[t]
 \centering 

   \includegraphics[width=3.4in]{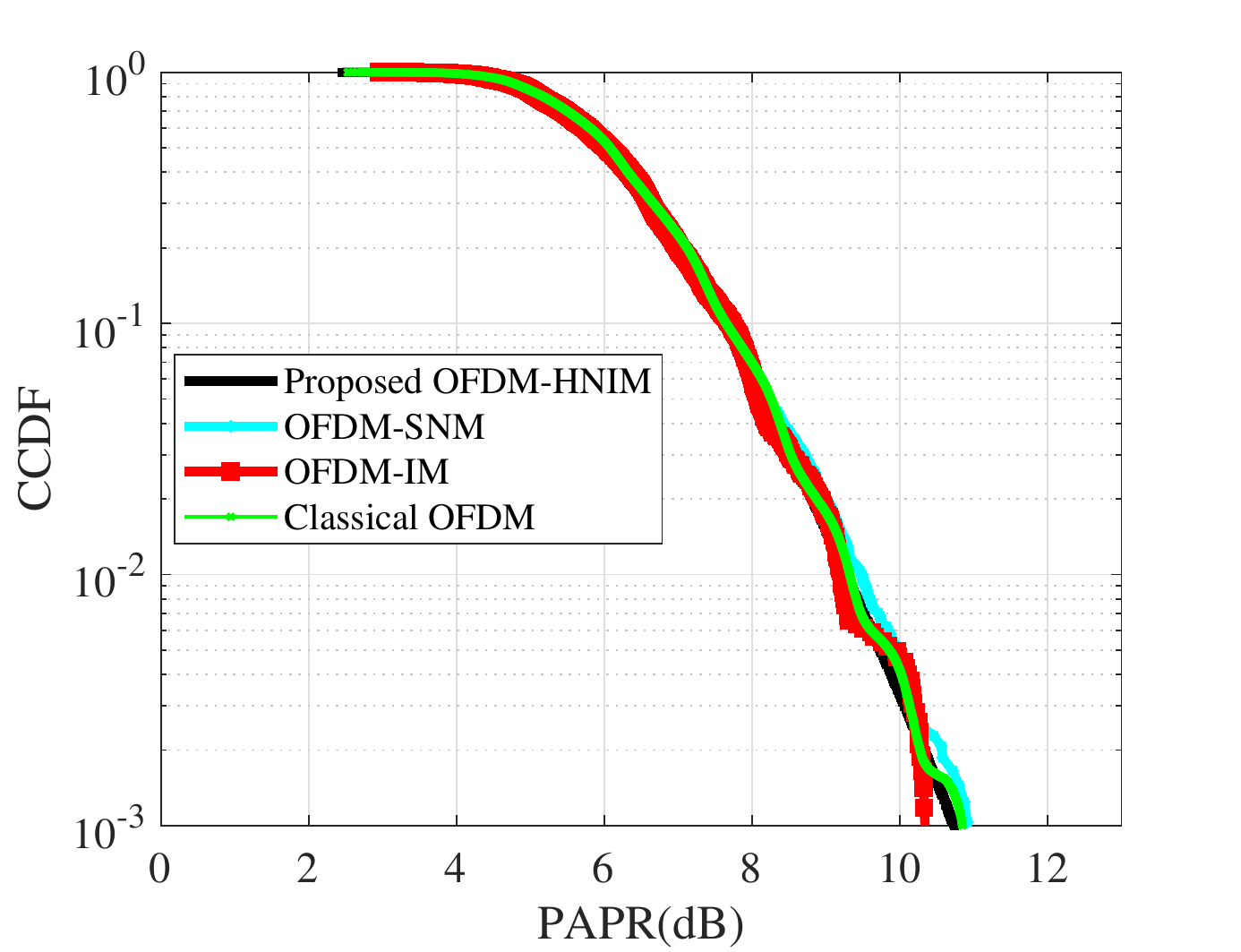}

 \caption{PAPR performances of the featured OFDM modulation options.}
 \label{paprcomp}
 \end{figure}

Fig. \ref{paprcomp} presents the complementary cumulative distribution functions (CCDFs) for the PAPR of the proposed OFDM-HNIM scheme compared to its counterparts OFDM-based modulation schemes. It can be observed from Fig. \ref{paprcomp} that the PAPR performances of these featured OFDM modulation options are high and almost the same \cite{Ishikawa2016,Jaradat2019}. The reason behind this high PAPR values featured in the proposed OFDM-HNIM scheme is the sparsity nature of the active subcarriers in the OFDM-HNIM which is also featured in the conventional OFDM-SNM and OFDM-IM schemes. These inherent features of having some inactive subcarriers in the OFDM-HNIM, OFDM-SNM, and OFDM-IM schemes could be exploited to reduce the PAPR by designing a proper mapping while ensuring that the PAPR in minimal levels. The PAPR reduction techniques proposed for conventional OFDM \cite{Rahmatallah2013} may not be directly applied to non-conventional OFDM schemes such as OFDM-HNIM scheme. The reason behind this is having unique features and characteristic for different OFDM modulation options \cite{Jaradat2019}. Therefore, we need a proper PAPR reduction method to reduce PAPR of the proposed OFDM-HNIM scheme, which is left as a future work.

\section{Conclusion} 
This paper proposes a new {energy and spectrally} efficient multi-carrier transmission scheme named hybrid number and index modulation that transmits additional information by number and index of the active subcarriers beside the conventional {signal constellation} symbols. The OFDM-SNM and OFDM-IM schemes are combined to harvest both of their advantages. The proposed hybrid scheme has low-complex transceiver which has outperformed its competitive schemes in terms of throughput performance under low order modulation {such as} BPSK, but it loses its dominance as modulation order increases. 
Moreover, Monte Carlo simulations have been conducted and the obtianed results have verified the analysis.
Due to having inactive subcarriers within the OFDM symbol featured in the proposed hybrid scheme, it could be used to lessen interference among subcarriers, and enhance {EE} by reducing PAPR. Furthermore, it is possible to design a proper power allocation technique to be integrated with the proposed hybrid scheme to improve its performance further.
\bibliographystyle{IEEEtran}

\end{document}